\documentclass[a4]{article}

\usepackage{arxiv}

\usepackage[utf8]{inputenc} %
\usepackage[T1]{fontenc}    %
\usepackage{hyperref}       %
\usepackage{url}            %
\usepackage{booktabs}       %
\usepackage{nicefrac}       %
\usepackage{microtype}      %
\usepackage{marginnote}

\expandafter\let\csname equation*\endcsname\relax

\expandafter\let\csname endequation*\endcsname\relax

\usepackage{mathtools}
\usepackage{amsfonts}       %
\usepackage{siunitx}

\usepackage{xstring}
\usepackage{catchfile}

\usepackage{xcolor}
\usepackage{graphicx}

\usepackage{multirow}

\usepackage[textsize=tiny]{todonotes}

\usepackage{url}
\usepackage{cleveref}

\let\vec\mathbf

\begin{document}

\title{Online Optimization of Stimulation Speed in an Auditory Brain-Computer Interface under Time Constraints}
\renewcommand{\shorttitle}{Online Optimization of Stimulation Speed in an Auditory BCI}
\author{Jan Sosulski \\
   University of Freiburg, Germany \\
   \texttt{jan.sosulski@blbt.uni-freiburg.de} \\
   \And
   David Hübner \\
   University of Freiburg, Germany \\
   \texttt{davhuebn@gmail.com} \\
   \And
   Aaron Klein \\
   University of Freiburg, Germany \\
   \texttt{kleinaa@cs.uni-freiburg.de} \\
   \And
   Michael Tangermann \\
   Radboud University, The Netherlands \\
   \texttt{michael.tangermann@donders.ru.nl} \\
 }

\maketitle

\begin{abstract}
The decoding of brain signals recorded via, e.g., an electroencephalogram, using machine learning is key to brain-computer interfaces (BCIs).
Stimulation parameters or other experimental settings of the BCI protocol typically are chosen according to the literature.
The decoding performance directly depends on the choice of parameters, as they influence the elicited brain signals and optimal parameters are subject-dependent.
Thus a fast and automated selection procedure for experimental parameters could greatly improve the usability of BCIs.\\
We evaluate a standalone random search and a combined Bayesian optimization with random search into a closed-loop auditory event-related potential protocol.
We aimed at finding the individually best stimulation speed---also known as stimulus onset asynchrony (SOA)---that maximizes the classification performance of a regularized linear discriminant analysis.
To make the Bayesian optimization feasible under noisy conditions and under the time pressure posed by an online BCI experiment, we first used offline simulations to initialize and constrain the internal optimization model.
Then we evaluated our approach online with 13 healthy subjects in the loop.\\
We could show that for eight out of thirteen subjects, the direct optimization of the classification performance succeeded to select the individually optimal SOA out of multiple evaluated SOA values.
Our data suggests, however, that subjects were influenced to very different degrees by the SOA: For the half of the subjects where SOA influenced the performance strongest, the approach that incorporates random search and Bayesian optimization succeeded for six out of seven subjects.
For the other half of the subjects, it succeeded for two out of six participants.\\
Our work proposes an approach to exploit the benefits of individualized experimental protocols and evaluated it in an auditory BCI.
When applied to other types of experimental parameters our approach could enhance the usability of BCI for different target groups---specifically if an individual disease progress may prevent the use of standard parameters.\\
 \end{abstract}


\section{Introduction}

Brain-computer interfaces (BCIs) are systems that decode individual signals of a user's brain to establish a closed-loop interaction.
Common applications are, for example, communication~\cite{farwell1988talking}, wheelchair control~\cite{wolpaw2012brain-computer,carlson2013brain-controlled}, motor rehabilitation after stroke~\cite{ang2013brain-computer} and non-clinical uses like gaming~\cite{krauledat2009playing,marshall2013games}.
To accomplish this, the decoding task consists of machine learning methods for classification or regression that are trained for each subject individually.
For instance, decoding the letter which a subject attends to on a screen \cite{farwell1988talking} or regressing the power of an electroencephalogram (EEG) component with task performance of the subject \cite{meinel2015eeg}.
The main challenge here is the intrinsically low signal-to-noise ratio (SNR) of multi-channel brain signal recordings, such as the EEG, and non-stationary feature distributions over time.
To establish closed-loop BCI applications, it is important to quickly, reliably and precisely decode the user's brain state.

The BCI's decoding performance depends on multiple factors, such as the choice of the experimental paradigm or the experimental parameters~\cite{amiri2013review}.
Many BCI applications use a sequence of external stimuli, which are presented to the subject.
Each stimulus is presented, for instance, in the auditory domain (e.g.~a sound or word) and elicits transient event-related potential (ERP) responses within a few hundred milliseconds after stimulus onset~\cite{farwell1988talking,schreuder2010new}, which can be recorded using EEG, magnetoencephalogram or invasive techniques.
Depending on the type and role of a stimulus, the elicited ERP responses have different spatio-temporal and amplitude characteristics.
They are used as features to classify the ERP in single-trial despite the high noise level.
The parameters which determine the stimulus presentation (e.g.~timing, speed, stimulus type and intensity) have an influence upon the single-trial discriminability of ERPs~\cite{hohne2012how,sugi2018improving,gonsalvez2007target-to-target,allison2006effects} and there is evidence, that optimal parameters are different for each subject~\cite{sugi2018improving}.
This difference could be exacerbated for, e.g., stroke victims, as the specific area that is lesioned is distinct for each patient.
Still, so far most stimulus presentation parameters---hereinafter referred to as stimulation parameters---are usually fixed, i.e.~no subject-specific parameters are used (e.g.~\cite{hohne2010two-dimensional,halder2015brain-controlled,lin2018novel}).

One classical paradigm that elicits ERPs is the auditory oddball task.
In this paradigm, a sequence of rare high-pitched target tones and frequent low-pitched non-target tones is presented to the subject, with the stimulation speed, i.e., the time between the onset of two successive tones being called the stimulus onset asynchrony (SOA).
The subject's task is to ignore the non-target sounds and to focus his/her attention onto target sounds.
As the evoked ERP responses differ between target and non-target stimuli, a classification model can be trained to decode the measured brain signals into target and non-target responses.
While this classification task has been investigated using different model classes, linear classification using a shrinkage regularized linear discriminant analysis (rLDA) delivers very good results despite its low capacity~\cite{blankertz2011single-trial}, which is why we also use it in this work.
For the auditory oddball paradigms, there is ample evidence, that the SOA influences both the shape of the ERPs and the discriminability of target/non-target stimulus classes.
For instance, very short SOAs produce small P300 amplitudes and usually yield bad classification performances, as target and non-target stimuli cannot be discriminated very well~\cite{hohne2012how,gonsalvez2007target-to-target}.
In this work, we aim to improve the accuracy of the machine learning classifier by finding subject-specific SOAs that maximize the amount of class-discriminative information in the data.
This optimization problem is treated as a one-dimensional black-box optimization problem.

One of the most successful frameworks for black box optimization problems, where no gradient information are available, is Bayesian optimization~\cite{mockus-tgo78a,jones-jgo98a}.
Due to its internal model, it often requires just a few samples of the objective function to approach the global optimum and hence, works particularly efficiently on functions that are expensive to evaluate.
In our case, each function evaluation is one trial with a given SOA where the subject performs the target detection task.
The involvement of a human subject in the optimization problem means that we can query the objective function only a limited amount of times as the subject, e.g., will get exhausted.

Bayesian optimization~\cite{snoek-nips12a} is one of the work horses to realize hyperparameter optimization (HPO)~\cite{feurer-automlbook18a} in machine learning.
It aims at optimizing the validation performance of a machine learning algorithm by tuning the hyperparameters which are specific for that algorithm.
In HPO, each evaluation of a set of hyperparameters requires to train and validate a machine learning model.
For neural networks, this can take several hours or even days.

The usage of Bayesian optimization in the context of BCI has so far been applied mainly to the underlying machine learning problem.
An exception is the work by Bashashati et al.~\cite{bashashati2016user-customized}.
The authors used Bayesian optimization one step earlier, i.e., to improve the parameters which govern the preprocessing of the data, feature preprocessing, and feature extraction steps.
Examples are frequency bands or time interval parameters, which shall be determined optimally for every subject.
This was applied to offline data.
Contrary to Bashashati and colleagues, we propose to profit from the benefits of automatic parameter optimization even earlier in the pipeline, i.e.,~at the level of the data generation during the execution of the BCI experiments.
This of course requires us to apply the Bayesian optimization online during the BCI experiment, while the human subject performs the task.

In this BCI setup we have to deal with a very unfavorable SNR, hard time constraints and are not able to perform any parallel function evaluations.
Each evaluation of the objective function corresponds to executing a trial of a BCI experiment, which, in our case, takes between 5 and 54 seconds.
Due to these durations, we can afford to use the computationally complex Bayesian optimization approach.
To tackle the aforementioned challenges, we evaluated two methods of optimizing the stimulation speed.
First, a random search~\cite{bergstra-jmlr12a} and second, a combination of random search and a Gaussian process based Bayesian optimization~\cite{snoek-nips12a} with additional domain-specific modifications.

So far, stimulation parameters are commonly taken from existing literature.
The few exploratory studies, which have optimized parameters made use of grid search strategies (e.g.~\cite{hohne2012how,sugi2018improving,jin2012targeting}).
Here, the stimulation parameters were measured repeatedly in order to find the optimal parameters for each subject.
However, due to the long duration needed to find the optimal parameters (e.g.~four hours in \cite{hohne2012how}) this is not feasible for single-session experiments.
For experiments with multiple sessions it is not clear yet, whether the subject-specific optimal stimulation parameters stay the same throughout each session.
In practice, the stimulation parameters that were optimal for most subjects are chosen in future experiments and for novel subjects, if no individual optimization is performed.
As a result, individual differences between subjects are neglected, resulting in suboptimal performances.

Additionally, in many of these studies the exact stimulation modalities and stimulation parameters are slightly different.
For instance, H\"ohne and Tangermann~\cite{hohne2012how} optimized the SOA with respect to information transfer rate (ITR) in an auditory oddball for a range of possible SOA values of 50\,ms to 1000\,ms.
They found that the optimal SOA for ITR ranges between 87 and 275\,ms depending on the subject, and for binary classification accuracy it ranged between 175 and 1000\,ms.
In a similar paradigm but using virtual sound sources with six different classes by Sugi et al.~\cite{sugi2018improving}, the authors tested SOAs between \SIrange{200}{1000}{\milli\second} and found 400 to \SI{500}{\milli\second} to be optimal for BCI utility (i.e., ITR) on average over all subjects, but they also found that some subjects performed very well with a 200\,ms SOA and therefore concluded that the ``appropriateness of a 200-ms SOA should be evaluated on a case by case basis''\cite{sugi2018improving}, reinforcing the need for subject-specific SOAs.
Finally, Jin et al.~\cite{jin2012targeting} optimized the target-to-interval (TTI) by varying the minimum number of non-target presentations between target stimuli.
When keeping the number of target and non-targets equal, changing the SOA also influences the TTI, hence making it a similar problem.

Lorenz et al. successfully applied Bayesian optimization to detect optimal experimental task parameters based on, e.g., fMRI recordings \cite{lorenz2016towards} or subjective behavioral scores \cite{lorenz2019efficiently}.
In their domain, they experience similar challenges as in our EEG-based BCI domain.
However, the underlying signal-to-noise ratio between the fMRI recordings and our EEG recordings cannot be accurately compared without a common metric that estimates the difficulty of the optimization problem.

\section{Online Bayesian optimization for BCI experiments}

\subsection{Subject-specific optimal stimulation parameters}
\label{sub:bo_in_bci}

In this work, we aim to optimize the SOA for subjects participating in an auditory oddball paradigm.
Our subjects were instructed to focus on the target tones (pitch 1000\,Hz) and ignore the non-target tones (pitch 500\,Hz).
Both kinds of tones were played via a speaker from a distance of 65\,cm from the subject for a duration of 40\,ms.
One trial of auditory oddball consisted of 15 target and 75 non-target stimuli in a pseudo-random order, such that there were always at least two non-target tones between two target tones.
Additionally, when dividing the 90 stimuli consecutively into 15 sets of six tones, there is always exactly one target tone in each set of six.
As the first few and the last few stimuli in a series are not independently and identically distributed~\cite{hubner2017challenging}, we omitted the first and last six epochs of each trial to avoid excessive outliers before further processing of the trial.
Therefore, we have 78 data points available per trial.

As the subject is in the loop during the SOA optimization, severe time restrictions apply.
For practical reasons, we imposed a time limit of 20 minutes to find the optimal SOA for a subject.
Furthermore, we restricted the SOA parameter space to be between \SIrange{60}{600}{\milli\second}, since shorter SOAs produced audio artifacts with our available hardware, and longer SOAs would introduce a number of drawbacks in the context of BCIs, e.g.~extensive trial durations.
The parameter space was sampled with a resolution of \SI{1}{\milli\second}, however, the effective resolution for the Gaussian process model depends on the estimated lengthscale of the kernel. 
As the stimulus duration was \SI{40}{\milli\second}, there was at least \SI{20}{\milli\second} pause between tones.

Our objective is to maximize the class discriminative information contained in the evoked target- vs.~non-target ERP responses.
In order to perform classification, we used the mean amplitude in the five time intervals: $T=\{[100, 170], [171, 230], [231, 300], [301, 410], [411, 500] \}$ in ms.
As we recorded with $N_c = 31$ channels, this yields a feature vector with $|T| \cdot N_c = 155$ dimensions for each epoch.
The discriminative information of the epochs in a trial is then estimated by the 13-fold cross-validated AUC score obtained on the feature vectors of one trial (corresponding to 78 data points) using an rLDA classifier with automatic shrinkage regularization of the estimated covariance matrix \cite{blankertz2011single-trial}.
Specifically, recording a single trial consists of a pseudo-randomized sequence of (after preprocessing) 13 target and 65 non-target stimulus presentations and takes---depending on the SOA---up to \SI{54}{\second}.

Clearly, the rLDA classifier will underperform due to the limited amount of training data compared to typical BCI studies.
However, as we want to explore the SOA space and as time restrictions for the full session apply, we aim to avoid aggregating data of multiple trials to obtain a more robust AUC estimate.
Limiting the AUC evaluation to the level of single trials is encouraged, as, for a given SOA the average performance of multiple rLDA classifiers trained on very small data subsets correlates strongly  with the performance of the rLDA trained on the complete data set. This was shown on similar data where the correlation between the average single trial AUC and the AUC on the complete dataset was reported as $\rho = 0.93$~\cite{sosulski2019extremely}.
Nevertheless, the resulting AUC value must be treated as a noisy and offset estimate of the performance that could be reached with a larger data set.
However, in the context of optimization, the AUC offset can be neglected.

\subsection{Bayesian optimization in an EEG-based auditory oddball BCI}

Given a function $f: \mathbb{R}^D \rightarrow \mathbb{R}$ based on some input $\mathbf{x} \in \mathcal{X} \subset \mathbb{R}^D$, Bayesian optimization (BO)~\cite{shahriari-ieee16a,jones-jgo98a} aims to find the global optimum $\mathbf{x}_\star \in \text{argmin}_{\mathbf{x} \in \mathcal{X}} f(\mathbf{x})$.
Thereby, we assume that $f$ can only be observed with noise: $y = f(\mathbf{x}) + \varepsilon_{noise}$ where $\varepsilon_{noise} \sim \mathcal{N}(0, \sigma_{noise}^2)$.
BO consists of two core components: a probabilistic model $p(f \mid \mathcal{D})$ that, based on some previously observed data $\mathcal{D} = \{(\mathbf{x}_1, y_1), \ldots, (\mathbf{x}_N, y_N) \}$, models the objective function, and a so-called \emph{acquisition function} $a_{p(f\mid\mathcal{D})}: \mathbb{R}^D \rightarrow \mathbb{R}$ which trades-off exploration and exploitation of the input space.
In each iteration, BO iterates the following steps: (1) select the next candidate point by optimizing the acquisition function $\mathbf{x}_{next} \in \text{argmax}_{\mathbf{x} \in \mathcal{X}} a(\mathbf{x})$ which is cheap to evaluate since it only relies on the model; (2) evaluate the objective function at $\mathbf{x}_{next}$ and obtain a new observation $y_{next}$; (3) augment the data $\mathcal{D} = \mathcal{D} \cup \{(\mathbf{x}_{next}, y_{next})\}$ and update the model.

The most frequently used models in BO are Gaussian processes (GP)~\cite{rasmussen-book06a} $f \sim \mathcal{GP}(m(\vec{x}), k(\vec{x},\vec{x}'))$ which are fully defined by a prior mean $m(\vec{x})$ (usually $m(\mathbf{x}) = 0 \, \, \forall \mathbf{x})$) and a kernel function $k(\mathbf{x}, \mathbf{x}')$ which measures the similarity between two points $\mathbf{x}$ and $ \mathbf{x}^{\prime}$ in the input space.
Following Snoek et al.~\cite{snoek-nips12a}, we use a Mat\'ern $\nicefrac{5}{2}$ kernel 
\begin{equation}
	k(\mathbf{x},\mathbf{x}^{\prime}) = \sigma_f^2 \left(1+\sqrt{5}d_\mathbf{\lambda}(\mathbf{x},\mathbf{x}')  +\nicefrac{5}{3}d^2_\mathbf{\lambda}(\mathbf{x},\mathbf{x}')\right)\operatorname{exp}({-\sqrt{5}d_\mathbf{\lambda}(\mathbf{x},\mathbf{x}')})
\end{equation}
where $d_\mathbf{\lambda}(\mathbf{x},\mathbf{x}')=(\mathbf{x}-\mathbf{x}')^T\operatorname{diag}(\mathbf{\lambda})(\mathbf{x}-\mathbf{x}')$ is the Mahalanobis distance.
Thereby, the signal variance  $\sigma_f^2$ and the lengthscales $\mathbf{\lambda}$ are considered to be hyperparameters of the GP.
Additionally, we use another hyperparameter $\hat{\sigma}_{noise}$ which is added to the diagonal of the covariance matrix to model the observation noise $\sigma_{noise}$.
To be more robust against the GP's hyperparameters $\theta = \left[\sigma_f, \mathbf{\lambda}, \hat{\sigma}_{noise}\right]$ we follow Snoek et al.~\cite{snoek-nips12a} and use the integrated acquisition function $   a_{p(f\mid\mathcal{D})}(\mathbf{x}) = \int a_{p(f\mid\mathcal{D}, \mathbf{\theta})}(\mathbf{x}) p(\mathbf{\theta} \mid \mathcal{D}) d\mathbf{\theta}$ by applying a Markov-Chain Monte-Carlo sampling of the marginal log-likelihood (see \cite{snoek-nips12a} for further details):
\begin{equation}\label{eq:marginal_ll}
  p(\mathbf{y} \mid \mathbf{X}, \mathbf{\theta}) = \mathcal{N}(\mathbf{y}| m, \Sigma_{\mathbf{\lambda}, \sigma_f} + \hat{\sigma}_{noise} \mathbb{I})
\end{equation}
As acquisition function, we used the upper confidence bound (UCB)~\cite{srninivas-icml10a} $a(\mathbf{x}) = \mu(\mathbf{x}) + \kappa \sigma(\mathbf{x})$ where $\mu(\mathbf{x})$, $\sigma(\mathbf{x})$ are the mean and standard deviation of the predictive distribution of the GP.
We do not use the common expected improvement acquisition function, since it relies on the currently best observed value, which is not reliable in case of high observation noise.
More complex information-theoretic acquisition functions, such as e.g.~entropy search~\cite{hennig2012entropy}, induce an computational overhead that is too expensive to be employed in our online experiments with strict time limitations.
For the experiments described in Section~\ref{sec:results} we used the Robust Bayesian optimization (RoBO) package~\cite{klein2017robo}.

Within the limited time horizon of 20 minutes, we expect to obtain only AUC estimates of 30-40 SOA values, each with considerable observation noise.
Therefore, we cannot afford to rely solely on the marginal likelihood for determining the GP-hyperparameters $\theta$.
Thus, in an offline simulation on ERP data of an auditory oddball under different SOAs  (see~\Cref{sub:offline_simulations}), we estimated expected bounds and initial values for $\theta$ across multiple subjects to warm-start the internal GP-hyperparameters optimization.
By employing this maximum a posteriori approach, we ensure that early iterations do not get stuck in local optima while still be able to learn subject-specific GP-hyperparameters once enough data is available.
We also determined $\kappa=2$ for the UCB acquisition function from this offline simulation.

To enforce additional exploration of BO especially during early iterations, we used an $\varepsilon$-greedy strategy which, with a probability of $\varepsilon$, samples a random data point from the parameter space (i.e., a random SOA) and with a probability of $1-\varepsilon$ evaluates the maximizer of the UCB acquisition function using DIRECT~\cite{jones-jgo01a} and uses the resulting SOA.
Our proposed approach therefore is a combination of classical BO combined with a random search.

The random sampling is inverse proportionally weighted to the SOA, such that SOAs which result in trials taking twice as long than trials with a shorter SOA, are half as probable.
The decay policy (cf.~Figure~2 in the supplementary) for $\varepsilon$ was also determined in offline simulations.
In the earliest iterations, we set $\varepsilon=1$ such that the first objective evaluations are made at random SOAs, which is helpful to build an accurate initial Gaussian process model.
Towards the end of the 20 minute time budget, $\varepsilon$ decays to 0, therefore only evaluating at SOAs suggested by the UCB acquisition function.

\subsection{Assessing feasibility of optimization with respect to noise}
\label{sub:onr}
The effectiveness of BO is compromised when the observation noise is substantially higher than the variance of the objective function.
Thus, it is necessary to relate the two factors to describe the difficulty of the optimization problem.

In signal processing, the notion of a SNR is well known and widely used.
However, the traditional SNR definition assumes time series and stationary signals.
If we assume noisy observations $y$ of the ground truth objective function $f$, we would like to quantify this amount of noise compared to the amount of change induced to $f$ when evaluating different parameters (SOAs in our case).
We thus propose to use the objective-to-noise ratio (ONR), which serves as an indicator of the difficulty of the optimization problem.
Small values of the ONR indicate that the observation noise is dominant compared to the overall change in the objective function.

The variance of the objective function can be approximated by densely sampling points in the parameter space.
However, since $f$ cannot be observed directly, we use the predictive mean $\mu$ of the fitted GP as our approximation, i.e.~$\hat{o}_\sigma = \sqrt{\mathrm{var}(\vec{m})}$, where $\vec{m} = [\mu(x_1), \ldots, \mu(x_N)]$ and $x_i \in X_s$, with $X_s$ = $[x_{min}, x_{min} + 1\cdot\nu, x_{min} + 2\cdot\nu, \ldots, x_{max}]$ and $x_{min}$ indicates the lower bound of the parameter space and $x_{max}$ the upper bound.
The step size $\nu$ should be as small as computationally feasible.
The observation noise $\sigma_{noise}$ is approximated by using the GP hyperparameter $\hat{\sigma}_{noise}$.
Note that now we can calculate this ONR as:
\begin{equation}
\label{eq:onr}
\text{ONR}\coloneqq{\hat{o}_\sigma}/{\hat{\sigma}_{noise}}.
\end{equation}
We expect low values for ONR, when there is either a large amount of noise or when the effect of changing the parameter has only little influence on the objective.

\subsection{Challenges in the domain of BCI} %
\label{ssub:challenges}

The EEG signals we measure contain brain activity as well as (undesired) non-neural activity.
As in our case we are interested in the event-related potentials, our signal of interest:
\begin{equation}\label{eq:nonstat}
	E = f(S) + T_n(\tau)
\end{equation}
is the synchronous brain response to the stimuli.
This signal $E$ depends on the stimulation parameters $S$ and is influenced by unknown non-stationarities $T_n$ depending on time $\tau$.
In our experimental setting, we hypothesize that the two major contributions to this non-stationarity are the decline of attention during the course of an EEG experiment's session and the habituation to the task~\cite{ravden1999on}.
Furthermore, slight changes in the paradigm between the optimization part and the validation of the experiment (cf.~\Cref{sub:experiments}) may also induce non-stationarity effects.
Using EEG however, we are only able to observe
\begin{equation}\label{eq:noise}
	\hat{E} = E + \varepsilon_{b} + \xi,
\end{equation}
which contains our signal of interest $E$ as well as additional background brain activity $\varepsilon_{b} \sim N(0, \sigma^2_{b})$ (with zero mean, as the data is high-pass filtered), which is unrelated to the task.
Additionally, in our signal we observe artifacts $\xi$ representing non-cortical electrical activity caused by, e.g.~eye blinks.

To get rid of $\xi$, typically an artifact removal step, i.e.~discarding sections of the EEG signal containing eye blinks etc., is performed.
To keep the amount of data constant in each trial, we omitted this artifact removal and instead tried to minimize the appearance of artifacts by instructing subjects to refrain from blinking and movement as much as possible.

\subsection{Offline simulations}
\label{sub:offline_simulations}

The offline simulation for determination of hyperparameter bounds for the Gaussian process is based on data collected in auditory oddball EEG experiments by Höhne and Tangermann~\cite{hohne2012how}.
On this data set $\mathcal{D}$ we used the same preprocessing and classification pipeline as described in Section~3.
The data set consisted of 14 subjects and 11 different SOA values that were tested in 16 trials each.

For the purposes of finding suitable hyperparameters for the Gaussian process, specifically the lengthscale $\lambda$ and the observation noise $\hat{\sigma}_{noise}$, we formed subsets $\mathcal{D}_{sub} \subset \mathcal{D}$ of our offline dataset for each subject.
From these subsets, we drew random trials and calculated single trial AUCs in cross-validation (cf.~Section~3).
Each trial corresponds to a specific SOA.
We used the SOA in combination with the obtained AUC values and fed them into a Gaussian process $\mathcal{GP}_{off}$.
This was done 500 times for every subject.
We used a region around the estimated hyperparameters of all the estimated $\mathcal{GP}_{off}$ and set them as hyperparameter bounds of the online Gaussian process $\mathcal{GP}$ used during the experiments.
The obtained bounds are:

\begin{itemize}
  \item $0.05 \leq \lambda \leq 1$
  \item $0.01 \leq \hat{\sigma}_{noise} \leq 1$
  \item $\sigma_f$ was not bounded
\end{itemize}

Note that in the used optimization framework, the input domain is normalized between 0 and 1.
The greatest benefit of these prior bounds is to restrict the lengthscale $\lambda$ as this forces the Gaussian process to ignore the extreme cases where a linear function of SOA explains the difference in performance in which $\lambda$ would grow very large.

\section{Experimental paradigm to evaluate optimization strategies in BCI} %
\label{sub:experiments}

\begin{figure}[ht]
      \centering
      \includegraphics[width=.85\linewidth]{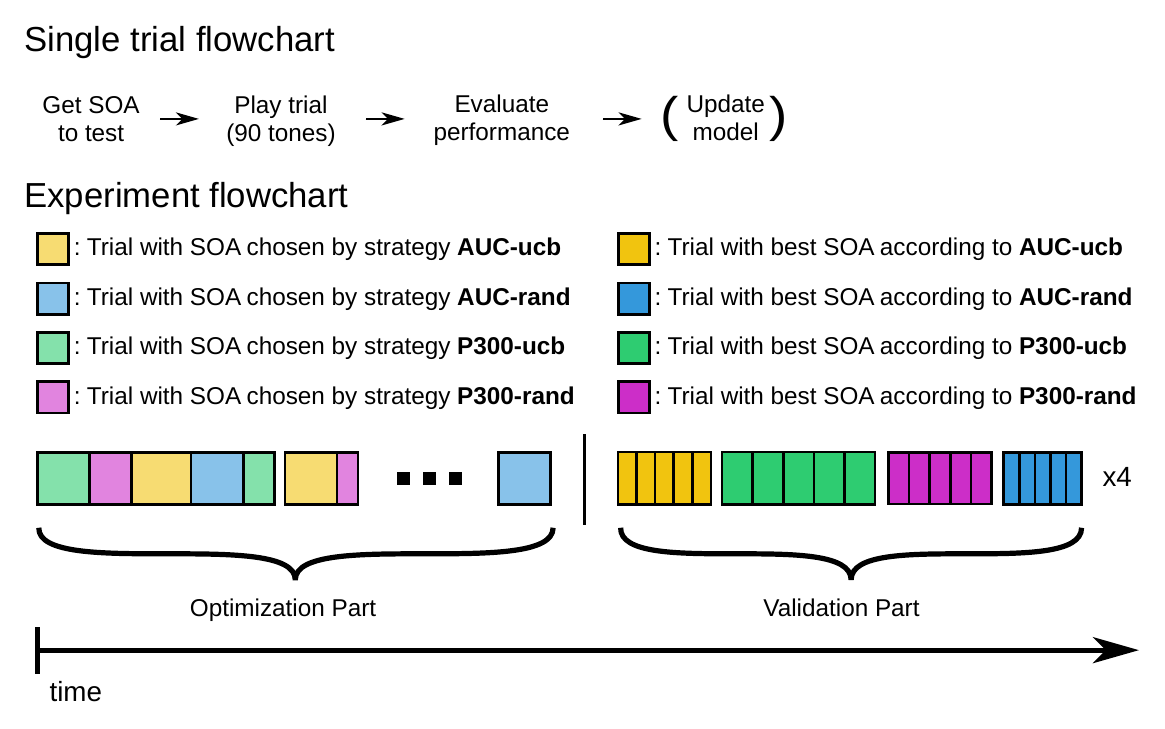}
      \caption{Schematic for the experimental setup.
      Top: before each trial, the SOA is set and after playing the 90 tones of a trial, performance (depending on the optimization target) is evaluated.
      During the first part (optimization part) of the experiment, the resulting performance is fed into the optimization model.
      Bottom: division into the two parts of the experiment.
      One block corresponds to a single trial.
      The block width indicates the trial length which depends on the SOA.
      Bottom left: in the optimization part, each optimization strategy change after each trial depending on the remaining time budget of each strategy.
      Bottom right: when each strategy has depleted its budget, the SOA validation part begins.
      The optimal SOAs as determined by each strategy are played consecutively in blocks of five trials until 20 trials are recorded for each SOA.}
      \label{fig:flow}
\end{figure}

Our online experiments were split into two parts.
This structure is shown in~\Cref{fig:flow}.
In the first part---the \textbf{optimization part}---we compared two BO approaches to optimize the optimal SOA, which only differ in the type of acquisition function.
Both were allocated an optimization time budget of 20 minutes each.
The first acquisition strategy (\textbf{ucb}) used the upper confidence bound acquisition function in combination with the previously explained decaying $\varepsilon$-random sampling.
The second one used a completely uninformed random acquisition strategy (\textbf{rand}) for the whole duration, similar to a random search.
To minimize the influence of non-stationarities within the optimization part which could affect our objective, the strategies were applied in an alternating order.
Additionally, subjects did not know which strategy suggested the current SOA.
Which strategy was allowed to suggest the first SOA was randomly chosen before beginning the experiment.
As the duration of an auditory oddball trial varies with the SOA, the strategies may have used up different amounts of their time budget after a trial.
Therefore, before starting a novel trial, we looked at the remaining time budgets.
For the next trial, the optimization strategy with the largest remaining time budget was allowed to choose the SOA that should be evaluated.
Note that in the optimization part of the experiment, nearly every successive trial was played at a different SOA.

As one optimization target we chose the aforementioned \textbf{AUC}.
Additionally, we applied both acquisition strategies also for maximizing an ERP feature, the \textbf{P300} amplitude, of target responses.
While this, in principle, is a neurologically motivated objective, the P300 response carries much of the discriminative information between target and non-target ERPs~\cite{hohne2012how}.
Thus, we expect SOAs resulting in strong target P300 responses to yield good classification performances, too.
The P300 latency was determined for each subject individually on the pooled data on the first few trials, before the first model-based SOA value was suggested.
Optimization strategies are named by combining the acquisition strategy with the optimization target, e.g.~\textbf{AUC-ucb} for optimization of the AUC using the described upper confidence bound acquisition strategy.

Since we do not have access to the ground truth of the best SOA values, we needed to validate the classification performances obtained for each SOA on novel data.
In this second \textbf{validation part} of the experiment, we evaluated the SOAs obtained by each optimization strategy in 20 trials each, therefore we obtain 20 AUC estimates for each SOA.
In most real-world BCI applications the SOA would not be changed at all during the full session.
In order to approximate this characteristic, we only change the SOA after every five trials during the validation part.
The advantage is, that the subject may be able to better adjust to an SOA, while as a potential drawback non-stationary effects (expressed by $T_n$ in Eq.
\ref{eq:nonstat}) may be introduced with respect to the optimization part.
However, we hypothesize that the effect of different SOAs exceeds non-stationary influences.

Hereinafter we report the mean of these 20 AUCs as the resulting performance for a given SOA.
Note that we did not train a single classifier on the pooled data of 20 trials in order to not change the objective (see~\Cref{sub:bo_in_bci}) between both parts of the experiment.
In the validation part, we always evaluated the fastest allowed SOA (\SI{60}{\milli\second}, \textbf{fixed60}).
We expect this SOA to perform poorly~\cite{hohne2012how} with classification performance close to chance level, but we still evaluate it in the validation, because it consumes very little experimental time and yields a considerable amount of data for post-hoc analysis.
Thus it serves as a worst case scenario regarding the expected classification performance.
Additionally, we evaluated the SOAs determined by each optimization strategy in the optimization part of the experiment.
As our maximum experimental time for the complete experiment (including EEG preparation time and breaks) was limited to four hours, we did not double-evaluate duplicate SOAs or highly similar SOAs.
The latter is defined as $\mathrm{SOA}_1\cdot1.1 > \mathrm{SOA}_2$ for $\mathrm{SOA}_1 \leq \mathrm{SOA}_2$.
In such cases we kept $\mathrm{SOA}_1$ only, except when $\mathrm{SOA}_2$ was determined by the strategy \textbf{AUC-ucb}.
In any of these cases, we finally assigned the same AUC values obtained in the validation to both strategies/SOA values.
However if all proposed SOA values were so long that the expected experimental time in the validation part would exceed four hours, we had to prune an additional SOA value.
It was chosen as the SOA which was closest to all others.

\begin{table}[ht]
\centering
\small
\caption{Summary of the used optimization strategies with their optimization target and the corresponding acquisition function.}

\begin{tabular}{l|ll}
\toprule
 \textbf{Strategy}  & \textbf{Target}  & \textbf{Acquisition} \\
 \midrule
 \textbf{AUC-ucb} & Single-trial AUC & Upper confidence bound\\
 \textbf{AUC-rand} & Single-trial AUC & Random sampling of SOAs\\
 \textbf{P300-ucb} & P300 amplitude & Upper confidence bound\\
 \textbf{P300-rand} & P300 amplitude & Random sampling of SOAs\\
\bottomrule
\end{tabular}

\label{tab:main_results}
\end{table}

In total, each subject yielded a data set for up to five (usually not unique) SOAs obtained by the five strategies: \textbf{AUC-ucb}, \textbf{AUC-rand}, \textbf{P300-ucb}, \textbf{P300-rand} and \textbf{fixed60}.
As the exact value for the SOA is different for each subject, we will hereinafter use the strategies for comparison purposes without exactly mentioning the obtained SOA value in milliseconds.

The 13 subjects (5 male, 8 female, mean age 22.7~$\pm$~1.64, range 20--26) gave written informed consent prior to participation in the experiment, and the experiments were approved by the ethics committee of the university medical center of Freiburg.
All experiments were performed according to the Declaration of Helsinki.
We recorded 31-channel EEG using a BrainProducts BrainAmp DC, with the passive Ag/AgCl electrodes placed according to the 10--10 system \cite{chatrian1985ten} for electrode positions.
The reference electrode was placed on the nose.
During both parts of the online experiment, we processed the EEG data in a short pause (approx.~\SI{8}{\second}) between successive trials.
Every five trials, the subject could take a longer pause if desired.
The recorded data was two-pass filtered between \SIrange{1.5}{40}{\hertz} using a fourth order inverse Chebyshev filter.
Then we created epochs from the continuous signal by segmenting the signal between \SIrange{-200}{1000}{\milli\second} around each stimulus.
The remaining epochs were corrected for baseline drifts, such that the sum of the measured EEG voltages in the interval of \SIrange{-200}{0}{\milli\second} was zero on each channel.

\section{Results of the online experiments} %
\label{sec:results}

\begin{figure*}[ht]
      \centering
      \includegraphics[width=\linewidth]{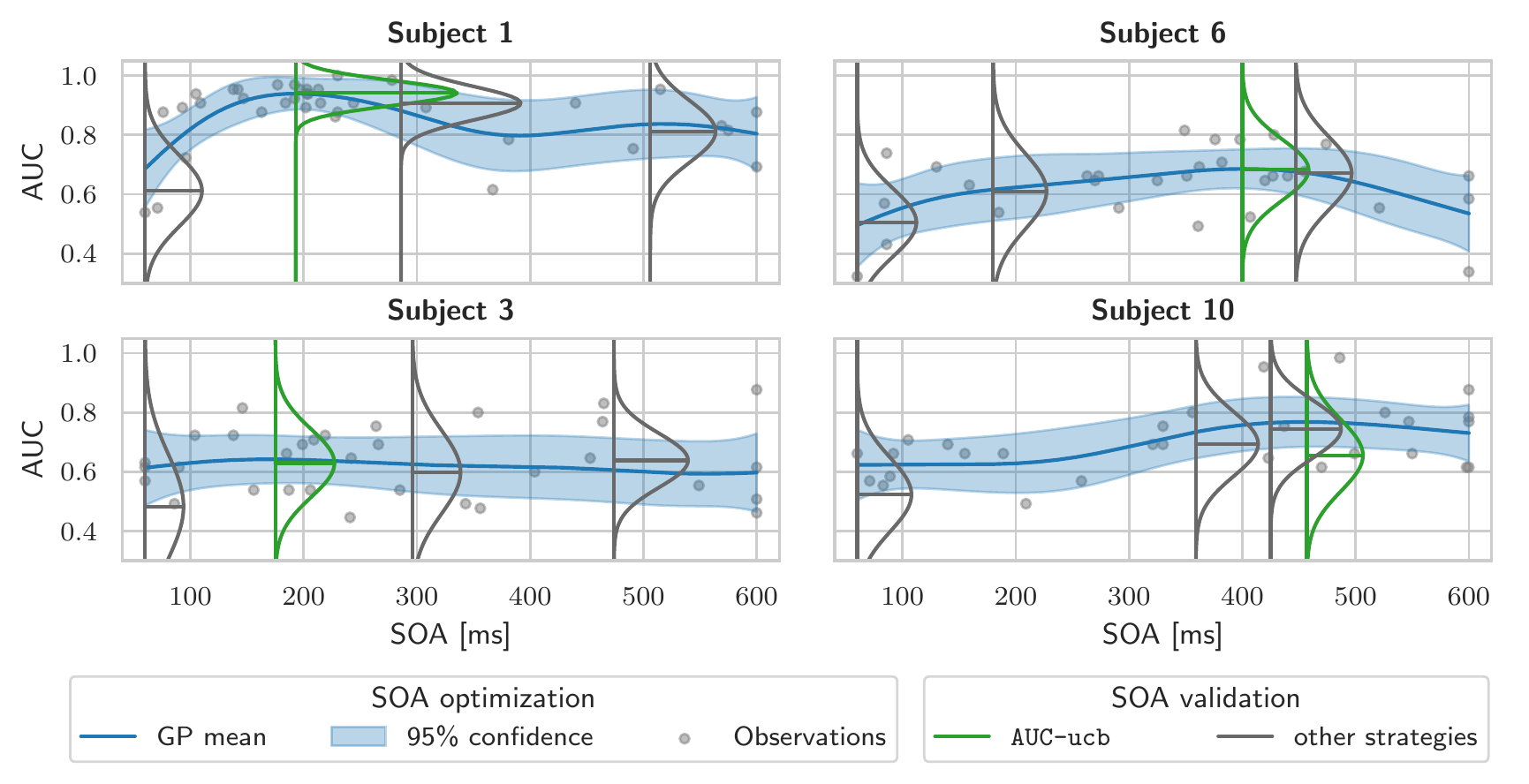}
      \caption{Final Gaussian processes and SOAs sampled by the end of the optimization process based on \textbf{AUC-ucb}.
      For the selected subjects (see main text), the gray circles indicate evaluated SOAs during the \textbf{optimization part} of the experiment.
      The resulting Gaussian process posterior mean and confidence intervals are visualized in blue.
      Green and gray vertical lines indicate which SOAs were selected by the strategies.
      The corresponding Gaussian distribution and horizontal mean thereof describe the result of evaluating these SOAs during the \textbf{validation part}.}
      \label{fig:four_subs}
\end{figure*}

At the end of the optimization part, each of the four strategies \textbf{AUC-ucb}, \textbf{AUC-rand}, \textbf{P300-ucb} and \textbf{P300-rand} had delivered an estimate for the optimal SOA.
Exemplarily, we show the GPs trained on the observed data and the sampled SOAs of \textbf{AUC-ucb} for four subjects in \Cref{fig:four_subs}.
Looking at the GP posterior mean (blue solid line) and its 95\,\% confidence interval (blue shaded area), we can see that SOAs around 200\,ms for subject 1 show the largest AUC values.
In contrast, the data of subject 3 was too noisy such that the no clearly optimal SOA range can be identified.
Note that the Gaussian processes for all subjects can be found in Figure~1 in the supplementary.

\begin{figure*}[ht]
      \centering
      \includegraphics[width=\linewidth]{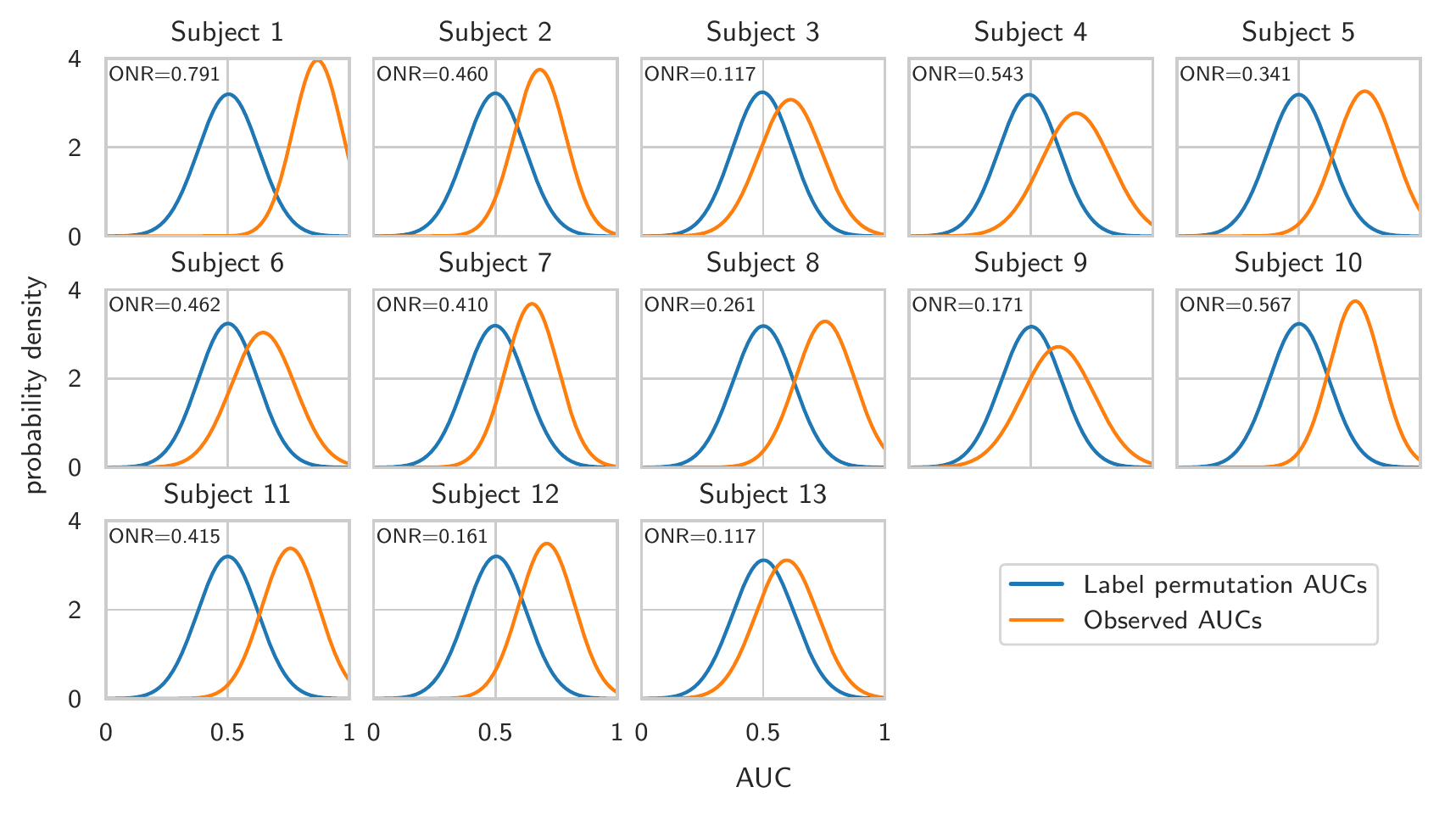}
      \caption{Results of label permutation testing for classification for all subjects.
      The blue curve indicates the distribution of AUC values when using random labels for the recorded trials.
      The distribution of the actually observed AUC values when using the true labels is indicated by the orange curve.}
      \label{fig:perm}
\end{figure*}

Given the small trial sizes, we additionally investigated how the single trial AUC values are distributed when using randomly permutated labels to investigate how much class-discriminative information is contained in the labeled data.
If the AUC distributions between the actual obtained AUCs and the AUCs on the permutated labels are very similar, there is not a lot of information contained in a single trial.
This would also show as a very low ONR.
For this, we used the trials recorded in the optimization part of the experiment and randomly permutated the labels of each trial and calculated the AUC value as described in \Cref{sub:bo_in_bci}.
As shown in \Cref{fig:perm}, for subjects 3, 9 and 13 the distribution of AUC values using randomly permutated labels overlaps largely with the actually observed AUCs using the true labels.
This indicates that for these subjects, there is hardly any information contained in the single trial AUC making the optimization process much harder.

\Cref{fig:four_subs} also shows the results of the validation part where we evaluated 20 trials from each proposed SOA.
Two example subjects for which the optimization worked well, are shown in the first row.
The SOAs (193\,ms for subject 1 and 400\,ms for subject 6) determined by the \textbf{AUC-ucb} strategy (indicated by the green vertical bars, located at the maximum value of the GP posterior mean curve) deliver the highest AUC values during the validation part.
Considering all four mean values of the validation SOAs, as indicated by the green and the three gray vertical lines and their distributions, we can see for both subjects that these means are in alignment with the estimated GP posterior mean (blue curve).

Two subjects, where our method showed suboptimal results, are shown in the second row of~\Cref{fig:four_subs}.
For subject 3, we argue that the small amount of variance in the estimated objective (the GP predictive mean) compared to the large amount of measurement noise, rendered the SOA optimization problem especially challenging given the 20 minute time budget.
This is supported by the observation that subject 3 showed one of the lowest ONRs among all subjects as shown in~\Cref{tab:complete_results}.
For subject 10, however, the ONR was quite large, but all AUCs obtained during the validation fell below the estimated GP posterior mean, where the SOA from \textbf{AUC-ucb} showed the greatest loss.
A possible reason for this mismatch between the performance in the optimization part and the validation part could be non-stationarities, such that the determined SOA was no longer optimal in the validation part of the experiment.
If that is the case, the hypothesis that the SOA influence overshadows non-stationarities is false for this subject.

To capture the influence of ONR on the success of our optimization goal, we sorted the subjects according to their ONR in~\Cref{tab:complete_results}.
In addition to the individual \textbf{ONR} values, also the average \textbf{AUC} values, obtained by cross-validation on data of the validation part, are reported.
These values were determined by applying the SOA which had been proposed by the \textbf{AUC-ucb} strategy at the end of the optimization part.
If this SOA turned out to deliver the highest AUC among all SOA values tested during the validation part, then the \textbf{Relative Performance} is 1.

For eight out of thirteen subjects, the SOA determined by the \textbf{AUC-ucb} strategy has found the best performing SOA of the validation part among the set of up to five tested SOAs.
We observed a failure to determine the best SOA in five subjects, which predominantly show a lower ONR (except for subject 10).
Different ONR values---as observed in our subject group---could be related to the phenomenon of BCI illiteracy~\cite{allison2010could}, which describes that brain signals of a certain percentage of the population cannot be decoded sufficiently well to e.g.~control a BCI application.

\Cref{tab:average} provides the grand average AUC across all subjects as obtained from the validation part of the experiment.
Bayesian optimization with an informed acquisition function (\textbf{AUC-ucb}) and a random acquisition function (\textbf{AUC-rand}) performed well.
Using a paired two-sided Wilcoxon signed rank test, the performance differences are significant between \textbf{AUC-ucb} and \textbf{P300-ucb} ($p=0.0027$) as well as the difference between \textbf{AUC-ucb} and \textbf{P300-rand} ($p=0.0349$).
Additionally, the difference to the \textbf{fixed60} condition is significant as well ($p < 0.001$).
The difference between \textbf{AUC-ucb} and \textbf{AUC-rand} was not significant ($p=0.6689$), even though the \textbf{AUC-ucb} yielded the optimal SOA for 8 subjects, whereas \textbf{AUC-rand} did so only for 7 subjects.
However, the number of subjects evaluated in this study is too low to conclude the advantage of either strategy over one another.
The given p-values were corrected for multiple testing using the Holm-Bonferroni correction.
While these results could be expected for \textbf{fixed60}, which had been included as a sanity check and as a lower performance boundary, the P300 strategies could be expected to deliver strong AUC performances, too (cf.~\Cref{sub:experiments}).
The exact found SOAs and their corresponding performances for each strategy are given in \Cref{tab:complete_results}.

\begin{table}[ht]
\centering
\caption{Comparison of the grand average performances across all subjects obtained from the validation part of the experiment via cross-validation using the five different SOA optimization strategies.
The standard error is computed based on the mean results for each subject grouped by strategy.}
\sisetup{table-format = -1.2}
\begin{tabular}[]{lS[table-format = 0.2]S[table-format = 0.2]S[table-format = 0.2]}%
\toprule
\textbf{Strategy} & \textbf{Mean AUC} & \textbf{Standard error} \\
\midrule
AUC-ucb & 0.701 & 0.027 \\
AUC-rand & 0.704 & 0.027 \\
P300-ucb & 0.670 & 0.016 \\
P300-rand & 0.681 & 0.025 \\
fixed60 & 0.517 & 0.013 \\
\bottomrule
\end{tabular}
\label{tab:average}
\end{table}

\begin{table}[ht]
\centering
\caption{Overview of results for the validation part of the experiment for all subjects.
The first column indicates the subject number.
Each cell belonging to a subject/strategy combination indicates the used stimulus onset asynchrony (SOA) as well as the mean AUC of the 20 measured trials and their standard deviation.
Note that for subject 4, the grouping of the SOAs (235 and 226) did not occur due to technical problems.
The number of subjects for which the strategy yielded the best performance is reported in the last row.}

\begin{tabular}{cccccccc}
\toprule
{Subject}  & {ONR} & {\texttt{AUC-ucb}}         & {\texttt{AUC-rand}}        & {\texttt{P300-ucb}}        & {\texttt{P300-rand}}       & {\texttt{fixed60}} \\
\midrule
\multirow{2}{*}{$1$} & \multirow{2}{*}{0.791} & $\mathbf{193}$             & $\mathbf{193}$             & $506$                      & $286$                      & $60$            \\
                     & & {$\mathbf{0.94 \pm 0.04}$} & {$\mathbf{0.94 \pm 0.04}$} & $0.81 \pm 0.11$            & $0.91 \pm 0.06$            & $0.61 \pm 0.12$ \\ \addlinespace[1.5mm]
\multirow{2}{*}{$10$}& \multirow{2}{*}{0.567} & $457$                      & $\mathbf{425}$             & $359$                      & $359$                      & $60$            \\
                     & & $0.66 \pm 0.12$            & {$\mathbf{0.74 \pm 0.10}$} & $0.69 \pm 0.11$            & $0.69 \pm 0.11$            & $0.52 \pm 0.13$ \\ \addlinespace[1.5mm]
\multirow{2}{*}{$4$} & \multirow{2}{*}{0.543} & $\mathbf{235}$             & $226$                      & $518$                      & $596$                      & $60$            \\
                     & & {$\mathbf{0.75 \pm 0.12}$} & $0.73 \pm 0.11$            & $0.67 \pm 0.13$            & $0.63 \pm 0.15$            & $0.48 \pm 0.11$ \\ \addlinespace[1.5mm]
\multirow{2}{*}{$6$} & \multirow{2}{*}{0.462} & $\mathbf{400}$             & $447$                      & $\mathbf{400}$             & $447$                      & $60$            \\
                     & & {$\mathbf{0.68 \pm 0.10}$} & $0.67 \pm 0.12$            & {$\mathbf{0.68 \pm 0.10}$} & $0.67 \pm 0.12$            & $0.50 \pm 0.12$ \\ \addlinespace[1.5mm]
\multirow{2}{*}{$2$} & \multirow{2}{*}{0.460} & $\mathbf{208}$             & $508$                      & $226$                      & $508$                      & $60$            \\
                     & & {$\mathbf{0.69 \pm 0.10}$} & $0.65 \pm 0.12$            & $0.67 \pm 0.11$            & $0.65 \pm 0.12$            & $0.51 \pm 0.13$ \\ \addlinespace[1.5mm]
\multirow{2}{*}{$11$}& \multirow{2}{*}{0.415} & $\mathbf{273}$             & $\mathbf{273}$             & $384$                      & $600$                      & $60$            \\
                     & & {$\mathbf{0.80 \pm 0.07}$} & {$\mathbf{0.80 \pm 0.07}$} & $0.72 \pm 0.13$            & $0.73 \pm 0.12$            & $0.52 \pm 0.11$ \\ \addlinespace[1.5mm]
\multirow{2}{*}{$7$} & \multirow{2}{*}{0.410} & $\mathbf{402}$             & $\mathbf{402}$             & $325$                      & $\mathbf{402}$             & $60$            \\
                     & & {$\mathbf{0.69 \pm 0.11}$} & {$\mathbf{0.69 \pm 0.11}$} & $0.59 \pm 0.11$            & {$\mathbf{0.69 \pm 0.11}$} & $0.50 \pm 0.13$ \\ \addlinespace[1.5mm]
\multirow{2}{*}{$5$} & \multirow{2}{*}{0.341} & $177$                      & $\mathbf{446}$             & $256$                      & $\mathbf{446}$             & $60$            \\
                     & & $0.70 \pm 0.11$            & {$\mathbf{0.74 \pm 0.11}$} & $0.71 \pm 0.11$            & {$\mathbf{0.74 \pm 0.11}$} & $0.57 \pm 0.15$ \\ \addlinespace[1.5mm]
\multirow{2}{*}{$8$} & \multirow{2}{*}{0.261} & $\mathbf{513}$             & $179$                      & $194$                      & $\mathbf{513}$             & $60$            \\
                     & & {$\mathbf{0.75 \pm 0.12}$} & $0.74 \pm 0.10$            & $0.69 \pm 0.11$            & {$\mathbf{0.75 \pm 0.12}$} & $0.57 \pm 0.12$ \\ \addlinespace[1.5mm]
\multirow{2}{*}{$9$} & \multirow{2}{*}{0.171} & $193$                      & $\mathbf{600}$             & $\mathbf{600}$             & $\mathbf{600}$             & $60$            \\
                     & & $0.56 \pm 0.13$            & {$\mathbf{0.60 \pm 0.07}$} & {$\mathbf{0.60 \pm 0.07}$} & {$\mathbf{0.60 \pm 0.07}$} & $0.46 \pm 0.14$ \\ \addlinespace[1.5mm]
\multirow{2}{*}{$12$}& \multirow{2}{*}{0.161} & $\mathbf{161}$             & $418$                      & $600$                      & $600$                      & $60$            \\
                     & & {$\mathbf{0.68 \pm 0.13}$} & $0.64 \pm 0.08$            & $0.62 \pm 0.10$            & $0.62 \pm 0.10$            & $0.51 \pm 0.13$ \\ \addlinespace[1.5mm]
\multirow{2}{*}{$3$} & \multirow{2}{*}{0.117} & $175$                      & $\mathbf{474}$             & $\mathbf{474}$             & $296$                      & $60$            \\
                     & & $0.63 \pm 0.12$            & {$\mathbf{0.64 \pm 0.09}$} & {$\mathbf{0.64 \pm 0.09}$} & $0.60 \pm 0.14$            & $0.48 \pm 0.18$ \\ \addlinespace[1.5mm]
\multirow{2}{*}{$13$}& \multirow{2}{*}{0.117} & $600$                      & $600$                      & $\mathbf{306}$             & $600$                      & $60$            \\
                     & & $0.57 \pm 0.14$            & $0.57 \pm 0.14$            & {$\mathbf{0.62 \pm 0.09}$} & $0.57 \pm 0.14$            & $0.46 \pm 0.14$ \\ \addlinespace[1.5mm]
\midrule
Best for & & \multirow{2}{*}{$\mathbf{8}$} & \multirow{2}{*}{$7$} & \multirow{2}{*}{$4$} & \multirow{2}{*}{$4$} & \multirow{2}{*}{$0$} \\
\# subjects & & & & \\
\bottomrule
\end{tabular}
\label{tab:complete_results}
\end{table}

\section{Discussion} %
\label{sec:discussion}

\textbf{Feasibility of the proposed strategies.}
Due to the high noise level of brain signals and limited amount of data, it was not clear whether a traditional random search or a random search combined with Bayesian optimization could be successfully applied.
The average results for thirteen subjects in~\Cref{tab:average} indicate that the optimization strategies based upon direct optimization of the AUC are feasible---as for no subject the performance using the optimized parameters degrades dramatically---and effectively lead to better performances compared to strategies which optimize only the class-discriminative P300 ERP.
In preliminary experiments, a non-domain specific Bayesian optimization approach produced unstable optimization courses, e.g.,~repeatedly testing at the boundaries of the parameter space or failing to recognize the extreme noise of the measurements by choosing too small kernel length-scales.
These issues were resolved by providing domain specific bounds and initial values to the GP hyperparameters as well as by using a random acquisition that does not rely on the model in early iterations when few measurements are available.

When looking at the obtained AUC values specifically of strategy \textbf{AUC-ucb} reported in~\Cref{tab:main_results}, we can see that this strategy found the best performing SOAs for most subjects.
However, we observed that with the low ONR of four out of the overall thirteen subjects, we may have touched the limits of the current optimization approaches.
Given the limited set of thirteen subjects and the substantial noise level in the EEG domain, we were unable to observe clear performance differences between the two acquisition functions \textbf{ucb} and \textbf{rand}.
This is less surprising if we keep in mind that our modified UCB strategy initially realizes a random search and that only for half of the trials (approx.~15-20) subject-specific SOAs were suggested by the model-based UCB acquisition function.
Hence, in the current study with only a single stimulation parameter and twenty minutes of optimization time, we could not determine whether the added Bayesian optimization is beneficial for the parameter search.

Given the current difficulty optimizing the noisy classification performance estimate of single trial AUCs, the prospect of optimizing multiple stimulation parameters at the same time has to be carefully evaluated.
However, this also depends on the used stimulation parameters and their influence on classification performance.
This suggests our preliminary analysis of the ONRs of stimulation parameter and optimization target combinations to determine feasibility of online optimization.
For problems which require the simultaneous optimization of a larger number of stimulation parameters, the search space grows exponentially.
In these cases, the \textbf{ucb} strategy that uses additional Bayesian optimization could have a distinct advantage compared to the plain random search.
The reason is, that an informed sampling strategy has a chance to detect a lower dimensional manifold containing good parameters, while the random strategy \textbf{rand} could fail to efficiently sample the large search space.

\textbf{Impact of noisy evaluations}
Especially subjects for which the ONR was small caused difficulties.
This is expected, because the optimization process practically operates only on random noise, if there is barely any influence of the SOA on AUC (cf.~subject 3 in~\Cref{fig:four_subs}).
An idea to alleviate the issues in future studies with low ONRs is to flexibly increase the fidelity of our data.
For subjects with extremely bad ONRs, one single trial of EEG data may not be sufficient to train an rLDA classifier that exceeds chance level.
In our case, we could train the rLDA on two successive trials with the same SOA instead of using only data of one single trial.
However, this would effectively double the evaluation times in the optimization part or result in fewer data points given the same time budget.
Additionally, the ONR metric could be used to inform a new kind of acquisition function, e.g., when the ONR values are too low reduce the confidence in the underlying Gaussian process and prefer the other exploration methods (in our search a plain random search).
The ONR metric can also be employed in any paradigm and model-based optimization method to quickly determine whether the current parameters that are optimized have enough influence on the features to make the optimization worthwhile.
A different approach to reduce the impact of the noisy evaluations is to employ classification methods that work especially well when few training data is available~\cite{sosulski2021improving}, which would reduce the noise in the estimation of the AUC.

\textbf{Violation of underlying assumptions.} Interestingly, the optimization was successful for particular subjects, e.g.,~subject 1, even though the data of this subject strongly violated some assumptions of BO: the AUC values approach a value of 1 for this subject (mind that subject 1 had the highest ONR), the assumption of normally distributed and homoscedastic noise is violated (cf.~\Cref{fig:four_subs}).
The observed robustness of the combined RS and BO approach under this obvious violation is promising regarding more challenging optimization objectives.
A neurotechnological application that has such challenging objectives is adaptive deep brain stimulation~\cite{beudel2016adaptive}, which tries to minimize stimulation side effects while maintaining the relief from symptoms.
However, even more closely related applications, such as a visual P300 speller~\cite{farwell1988talking}, where the visual ERPs tend to be easier to classify, could consistently yield AUC values close to 1.
If in these cases the violations make a classical Gaussian process unfeasible, we would consider alternative approaches that have been proposed which are capable to model heteroscedastic noise~\cite{kersting2007most}, non-normal noise distributions~\cite{vanhatalo2009gaussian} or use warped Gaussian processes~\cite{snelson2004warped} which are able to handle non-linear and non-normally distributed optimization targets.

\section{Conclusion} %
\label{sec:conclusion}

We translated a plain random search and a combined random search and Bayesian optimization approach, commonly used for hyperparameter optimization in machine learning, into a physical BCI experimental scenario, which is characterized by strict time limitations for experimenting and large noise in the observed data.

In the context of neurotechnological systems, our proposed approach is promising for designing new EEG-based paradigms for which prior knowledge about good experimental parameters (e.g.~parameters related to timing, stimulation) is lacking, or if good parameters are highly variable between subjects.
We specifically foresee BCI-supported rehabilitation training, e.g.,~after a stroke~\cite{ang2013brain-computer}, as a valuable application regime for the proposed method.
Here patient-specific trainings are of particular importance, since the brain structures affected by stroke and the resulting deficits vary from patient to patient.

Our results indicate that a subject-specific optimization of the SOA in an auditory oddball experiment is feasible, even with extremely noisy objective evaluations of non-invasive brain signal recordings.
We expect that a notable performance difference between an upper confidence bound acquisition function and an uninformed random acquisition function could be obtained for experiments, that require to determine more than a single parameter only, as the sample efficiency of Bayesian optimization~\cite{rai2019using} is more impactful in high dimensions.
In these situations, however, the practical limits of recording time available within a single experimental session could be exceeded, which should be evaluated in further work.
For a low-dimensional parameter problem as the one analyzed, we see the possibility to make more efficient use of the existing data points, e.g., using multiple trials close in the parameter space to acquire an estimate of higher fidelity, in order to outperform uninformed acquisition strategies.

In summary, we think that the neurotechnology community can profit from prior work on hyperparameter optimization generated in the field of machine learning.
However, we clearly see the need to conduct research into domain-specific adaptations of Bayesian optimization approaches to efficiently apply them for the purpose of optimizing experimental parameters.

\section*{Acknowledgment}
\label{sec:acknowledgment}

Funding: This work was supported by the Cluster of Excellence BrainLinks-BrainTools funded by the German Research Foundation (DFG) [grant number EXC 1086], by the DFG project SuitAble [grant number TA 1258/1-1] and by the state of Baden-W\"{u}rttemberg, Germany, through bwHPC and the German Research Foundation (DFG) [grant number INST 39/963-1 FUGG].

\section*{Conflicts of interest}

The authors declare they have no conflicts of interest.

\section*{Data availability statement}
The raw recorded EEG data is available online under the URL given below, the recorded classification performances for each trial of each subject and SOA is available in the supplementary.

\noindent\url{https://freidok.uni-freiburg.de/data/154576}

\bibliographystyle{unsrt}
\bibliography{bib/master,bib/lib_aaron}

\clearpage

\end{document}


\title{Supplementary Material for: Bayesian Optimization of Stimulation Parameters in Brain-Computer Interfaces under Time Constraints and Extreme Noise}

\author{Anonymous Authors}

\maketitle

\appendix

\section{Full results for part 1 (SOA optimization)}

\begin{figure}[b]
      \centering
      \includegraphics[width=\linewidth]{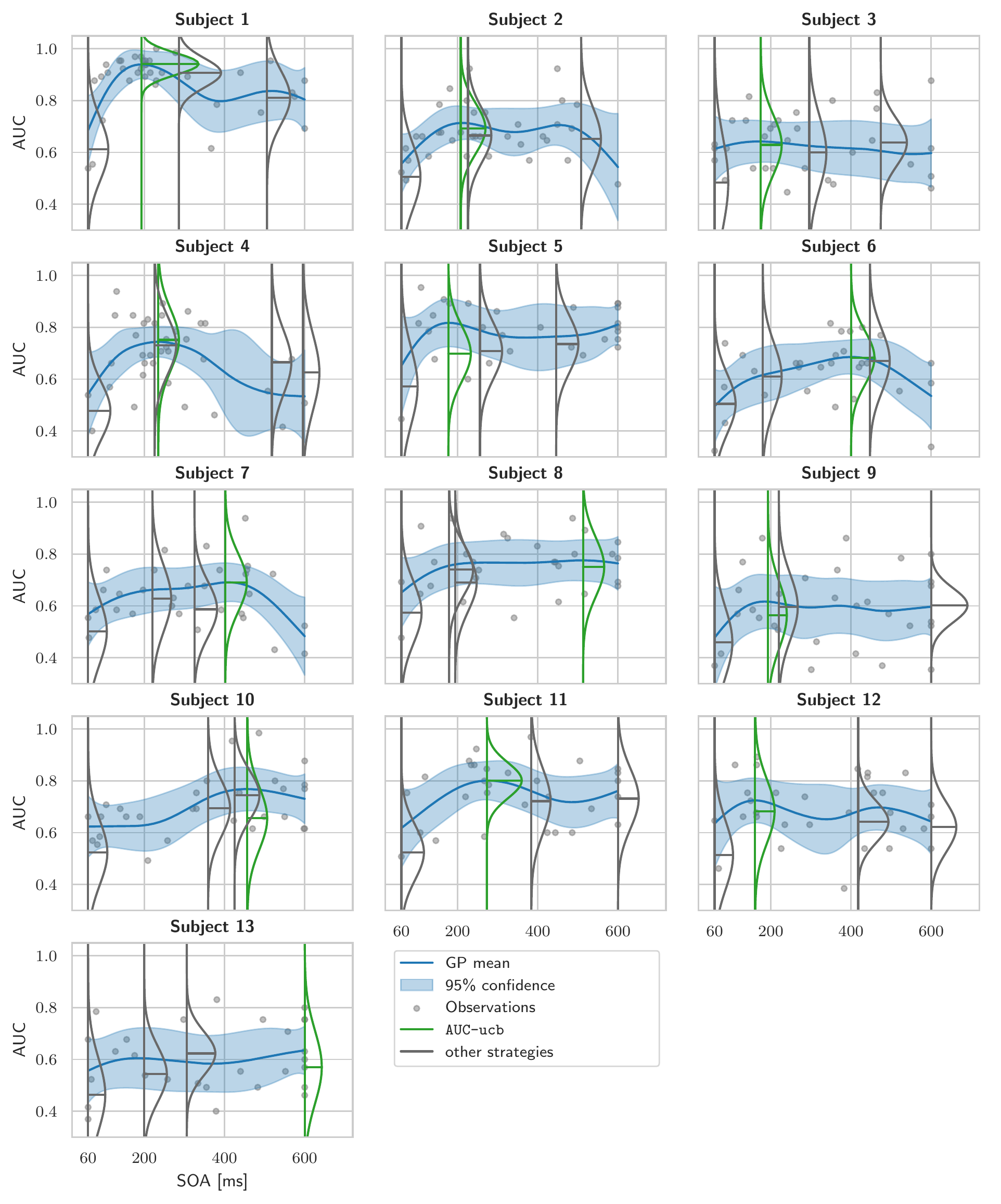}
      \caption{Final Gaussian processes and SOAs sampled by the end of part 1 using strategy \texttt{AUC-ucb}. The gray circles indicate evaluated SOAs, while the resulting GP posterior mean and confidence intervals are visualized in blue. Green and gray vertical lines indicate which SOAs were selected by the strategies. The corresponding Gaussian distribution and horizontal mean thereof describe the result of evaluating this SOA during part 2.}
      \label{fig:full}
\end{figure}

\Cref{fig:full} is generated using the data in \texttt{part\_1\_optimization\_auc\_ucb\_data.csv} and \texttt{part\_2\_validation\_data.csv} and shows the final Gaussian processes of the \texttt{AUC-ucb} strategy, as well as the distribution of the AUCs obtained in part 2 of the experiment.
The recorded data (subject number, trial number, evaluated SOA, AUC) in \texttt{part\_1\_optimization\_auc\_ucb\_data.csv} relates to the \texttt{AUC-ucb} strategy.

\section{Decay policy for epsilon}

\begin{figure}[b]
      \centering
      \includegraphics[width=.7\linewidth]{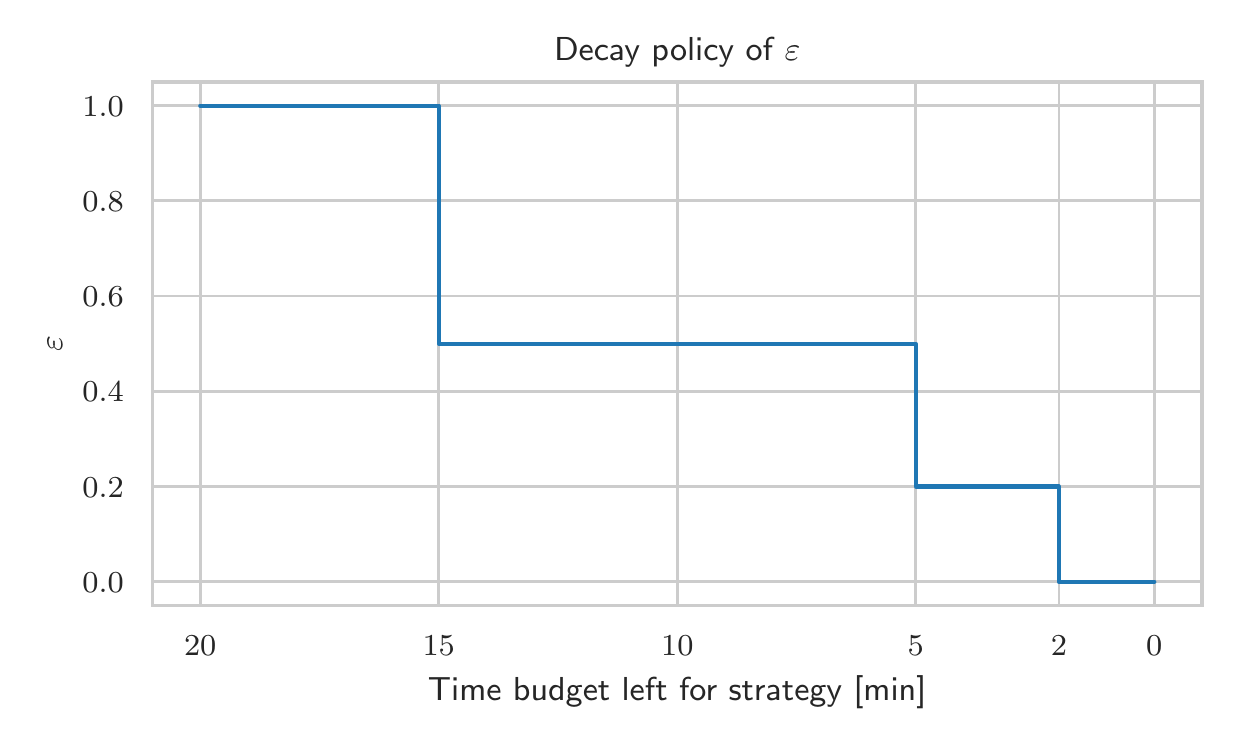}
      \caption{Decay policy for $\varepsilon$ for UCB strategies used in this paper. In the first moments of the experiment, SOAs were chosen randomly, as no reliable underlying model can be built with only few very noisy observations.}
      \label{fig:decay}
\end{figure}

The used decay policy for the $\varepsilon$ parameter is shown in~\Cref{fig:decay}.

\section{Full results for part 2 (SOA validation)}

The complete results can be found in the attached \texttt{part\_2\_validation\_data.csv}. Note that the strategies in this csv-file are identified by a number as shown in \Cref{tab:matching}:

\begin{table}[h!]
\centering
\caption{Matching of numbers to strategies.}
\begin{tabular}{ccccccc}
\toprule
{\texttt{AUC-ucb}} & {\texttt{AUC-rand}} & {\texttt{P300-ucb}} & {\texttt{P300-rand}} & {\texttt{fixed60}} & {\texttt{manual}} \\
\midrule
1 & 2 & 3 & 4 & 5 & 6 \\
\bottomrule
\end{tabular}
\label{tab:matching}
\end{table}

In Table 4 of the main paper, the contents of this csv file are shown averaged for each subject/strategy combination. An SOA according to the strategy \texttt{manual} was only chosen, when more than one SOA was dropped due to similarity. The resulting trials and mean AUC values were only used for ERP analysis and not in the context of any analysis in this paper, except for visualization purposes in Figure 2 of the main paper.